\documentclass[aps,pre,twocolumn,showpacs,showkeys,superscriptaddress]{revtex4}
\usepackage{amsmath}
\usepackage{bm}
\usepackage{graphicx}
\usepackage{hyperref}

\begin{document}           

\leftline{Draft version of Dec. 2011, for JPP or PoP or ...}

\title{Linear Kinetic Coupling of Firehose (KAW) and Mirror Mode}
\author{Hua-sheng XIE\footnote{Email: huashengxie@gmail.com}}
\affiliation{Institute for Fusion Theory and Simulation, Zhejiang
University, Hangzhou, 310027, PRC}
\author{Liu CHEN\footnote{Email: liuchen@uci.edu}}
\affiliation{Institute for Fusion Theory and Simulation, Zhejiang
University, Hangzhou, 310027, PRC} \affiliation{Department of
Physics and Astronomy, University of California, Irvine, CA
92697-4575, USA}
\date{\today}

\begin{abstract}
A general gyrokinetic dispersion relation is gotten and is applied
to analysis linear kinetic coupling of anisotropic firehose (or,
kinetic Alfv\'en wave) and mirror mode. Nyquist stability analysis
is also given.
\end{abstract}

\pacs{...}


\maketitle

\section{Introduction}\label{sec:intro}\suppressfloats
Firehose (FH) and Mirror Mode (MM) instabilities are both pressure
anisotropic instabilities, which mainly happen in high beta plasma,
and have many applications in space and astrophysical physics. One
(FH) happens when parallel pressure exceeding perpendicular one, and
the other (MM) happens when perpendicular pressure exceeding
parallel one. Both of these two modes have been discussed in
literatures by many authors. The basic properties of them can also
be found in textbooks or monographs, see e.g., \cite{Hasegawa1975a}
and \cite{Treumann1997}.

Mirror mode (MM) instability was identified and discussed since
\cite{Rudakov1961}, where the MHD theory is used, which is only
suitable for long wavelength limit. \cite{Tajiri1967} gives a
kinetic description, and shows that this mode is not a simply fluid
instability. \cite{Hasegawa1969} put forward to both the effects of
finite Larmor radius (FLR) and nonuniform, i.e., drift-mirror mode.
\cite{Southwood1993} gives a discussion of the physical mechanism of
the linear mirror instability in cold electron temperature limit.
Recently, series papers by Pokhotelov \textit{et al} addressed many
details of MM via both fluid theory and kinetic theory, such as
finite electron temperature effects \cite{Pokhotelov2000},
non-Maxwellian velocity distribution \cite{Pokhotelov2002}, finite
ion-Larmor radius wavelengths \cite{Pokhotelov2004, Pokhotelov2006}.
Gyrokinetic theory and simulation are introduced to discuss linear
and nonlinear MM by Qu \textit{et al}, \cite{Qu2007, Qu2008,
Qu2008a}.

Firehose instability is another well-known plasma instability, which
is in fact from the same branch of the kinetic Alfven wave (KAW),
hence, we will not distinguish them in the following sections. When
isotropic but with FLR, we get KAW \cite{Hasegawa1975,
Hasegawa1976}; while, when anisotropic but using long wavelength
limit, we get the classical FH \cite{Rosenbluth1956,
Chandrasekhar1958, Parker1958}. It is found by \cite{Yoon1993} (also
a review paper \cite{Yoon1995}) that a significant new effect which
had been neglected in the past will be brought in when allowing the
ion gyroradius to be finite. \cite{Chen2010} extends \cite{Yoon1993}
to discuss electron temperature anisotropy effects.

FH and MM have been discussed in almost the same approaches due to
their similarities, e.g., CGL fluid, kinetic corrections. However,
most the analytical solutions are reduced to just including one of
them, and then, there are few discussions of coupling effects. For
examples, when mirror mode is unstable, where is the FH (KAW)
solution? Can FH be also excited to unstable? Can FH bring MM from a
pure damping/growth mode to with real frequency?
\cite{Schekochihin2008} just discusses the nonlinear growth of FH
and mirror fluctuations, mainly via simulation. \cite{Duhau1985} has
discussed the coupling indeed, but uses the fluid theory then not
general. The gyrokinetic treatment of the stability of coupled
Alfv\'en and drift mirror modes in non-uniform plasma can be found
in \cite{Klimushkin2011, Klimushkin2012}, however, where FLR effects
are neglected.

Here, we talk about linear coupling effects to answer the above
questions. Firstly, a general anisotropic (bi-Maxwellian) 3-by-3
dispersion relation matrix is derived in gyrokinetic framework
(then, with FLR effects but only for ${k_\parallel }/{k_ \bot } \ll
1$) in Sec.\ref{sec:dr}. In Sec.\ref{sec:fhmm}, we show the
gyrokinetic dispersion relation can reproduce the FH and MM
solutions. In Sec.\ref{sec:coupling}, using only cold electron
assumption ${T_e}/{T_i} \ll 1$, the matrix is reduced to 2-by-2 for
discussion the coupling of FH and MM. The dispersion relation is
solved both analytical and numerical, and is also compared with the
full-kinetic code WHAMP (Waves in Homogeneous Anisotropic
Multicomponent Magnetized Plasma, \cite{Ronnmark1983}). General
stability properties for arbitrary Larmor radius at the $({\beta
_\parallel },{\beta _ \bot })$  plane are made clearly from both
analytical and numerical solutions and are also confirmed by Nyquist
analysis in Sec.\ref{sec:nyquist}. A summary is drawn in the last
section. Appendix gives some detailed discussions of the gyrokinetic
dispersion matrix.

\section{Gyrokinetic Dispersion Relation Matrix}\label{sec:dr}\suppressfloats
Here, we use the gyrokinetic (GK) theory \cite{Brizard2007} instead
of the full kinetic (FK) theory to calculate the general dispersion
relation matrix, which excludes the high frequency (${\Omega _{ci}}$
) modes automatically. The linear GK ordering is
\begin{enumerate}
\item Small amplitude: $\delta f/{F_0} \sim e\delta \varphi /T \sim \delta B/{B_0} \sim \delta  \ll
1$;
\item Low frequency: $\omega /{\Omega _i} \sim \delta $;
\item Anisotropic: ${k_\parallel }/{k_ \bot }\sim{\rho _i}/{L_0} \sim \delta,~{k_ \bot }{\rho _i} \sim
1$.
\end{enumerate}

For detailed discussion of the gyrokinetic assumptions, one can
refer to \cite{Brizard2007} or \cite{Howes2006}. We comment here,
the same ordering $\delta$  used here for all small variables is
just for convenience, and the third assumption ${k_ \bot }{\rho _i}
\sim 1$ is not a must.

Electrons and ions are assumed both bi-Maxwellian distribution,
\begin{equation}\label{eq:dis}
{F_0} = {1 \over {{\pi ^{3/2}}\alpha _ \bot ^2{\alpha _\parallel
}}}\exp \left( { - {{v_ \bot ^2} \over {\alpha _ \bot ^2}} -
{{v_\parallel ^2} \over {\alpha _\parallel ^2}}} \right),~{\alpha _{
\bot ,\parallel }} = {\left( {{{2{T_{ \bot ,\parallel }}} \over m}}
\right)^{1/2}}.
\end{equation}

Using the framework of \cite{Chen1991}, we write the linearized
equations of quasi-neutrality condition, vorticity equation (or
parallel Ampere's law) and perpendicular Ampere's law to matrix form
\begin{equation}\label{eq:gkeq}
\mathord{\buildrel{\lower3pt\hbox{$\scriptscriptstyle\leftrightarrow$}}\over
 C} \left[ \begin{array}{c}
  \delta {\phi _\parallel } \cr
  \delta \psi  \cr
  {{\delta {B_\parallel }} \over {{B_0}}}{{{T_{ \bot i}}} \over {{q_i}}} \end{array}  \right] \equiv \left(
  {\begin{array}{ccc}
   {{c_{SS}}} & {{c_{AS}}} & {{c_{MS}}}  \cr
   {{c_{SA}}} & {{c_{AA}}} & {{c_{MA}}}  \cr
   {{c_{SM}}} & {{c_{AM}}} & {{c_{MM}}}  \cr
 \end{array} } \right)\left[ \begin{array}{c}
  \delta {\phi _\parallel } \cr
  \delta \psi  \cr
  {{\delta {B_\parallel }} \over {{B_0}}}{{{T_{ \bot i}}} \over {{q_i}}} \end{array}  \right] = 0.
\end{equation}
Here, Coulomb gauge $\nabla  \cdot \mathbf{A} = 0$ is used. $\delta
\phi$ is the perturbed electrostatic potential, $\delta \psi  =
\delta {A_\parallel } \cdot \omega /c{k_\parallel }$ is a quantity
related to the parallel component of the perturbed magnetic vector
potential, and $\delta {B_\parallel }$ is the parallel component of
the perturbed magnetic field. The final dispersion matrix can be
gotten as,
\begin{eqnarray}\label{eq:drmatrix}
\mathord{\buildrel{\lower3pt\hbox{$\scriptscriptstyle\leftrightarrow$}}\over
 C}  \equiv \left( {\begin{array}{ccc}
   {{c_{SS}}} & {{c_{AS}}} & {{c_{MS}}}  \cr
   {{c_{SA}}} & {{c_{AA}}} & {{c_{MA}}}  \cr
   {{c_{SM}}} & {{c_{AM}}} & {{c_{MM}}}  \cr
 \end{array} } \right)\nonumber \\
 = \left( {\begin{array}{ccc}
   { - {\lambda _1} + {\lambda _2}} & { - {\eta _i}{\lambda _5}} & { - {\lambda _3} - {\lambda _4}}  \cr
   { - {\eta _i}{\lambda _5}} & {{{{\eta _i}{b_i}} \over {{{\bar \omega }^2}}}(1 + \Delta ) - {\eta _i}{\lambda _5}} & {{\eta _i}{\lambda _6}}  \cr
   { - {\lambda _3} - {\lambda _4}} & {{\eta _i}{\lambda _6}} & {{{2{\eta _i}} \over {{\beta _{ \bot i}}}} - {\lambda _7} - {\lambda _8}}  \cr
 \end{array} } \right)
\end{eqnarray}
where,
\begin{eqnarray*}\label{eq:drlambda}
\Delta  = {{{\beta _{ \bot i}}} \over {2{b_i}}}\left( {1 - {\eta
_i}} \right)\left[ {1 - {\Gamma _0}({b_i})} \right] + {{{\beta _{
\bot e}}} \over {2{b_e}}}\left( {1 - {\eta _e}} \right)\left[ {1 -
{\Gamma _0}({b_e})} \right],\\
{\lambda _1} = \left[ {1 + {\xi _i}Z({\xi _i}){\Gamma _0}({b_i})}
\right] + \tau \left[ {1 + {\xi _e}Z({\xi _e}){\Gamma _0}({b_e})}
\right], \\
{\lambda _2} = \left( {1 - {\eta _i}} \right)\left[ {1 - {\Gamma
_0}({b_i})} \right] + \tau \left( {1 - {\eta _e}} \right)\left[ {1 -
{\Gamma _0}({b_e})} \right],\\
{\lambda _3} = \left( {1 - {\eta _i}} \right){\Gamma _1}({b_i}) -
{{{\eta _i}} \over {{\eta _e}}}\left( {1 - {\eta _e}} \right){\Gamma
_1}({b_e}),\\
{\lambda _4} = {\xi _i}Z({\xi _i}){\Gamma _1}({b_i}) - {{{\eta _i}}
\over {{\eta _e}}}{\xi _e}Z({\xi _e}){\Gamma _1}({b_e}), \\
{\lambda _5} = \left[ {1 - {\Gamma _0}({b_i})} \right] + {\tau _
\bot }\left[ {1 - {\Gamma _0}({b_e})} \right], \\
{\lambda _6} = {\Gamma _1}({b_i}) - {\Gamma _1}({b_e}), \\
{\lambda _7} = (1 - {\eta _i}){\Gamma _2}({b_i}) + {{{\eta _i}}
\over {{\eta _e}{\tau _ \bot }}}(1 - {\eta _e}){\Gamma _2}({b_e}),
\\
{\lambda _8} = {\xi _i}Z({\xi _i}){\Gamma _2}({b_i}) + {{{\eta _i}}
\over {{\eta _e}{\tau _ \bot }}}{\xi _e}Z({\xi _e}){\Gamma
  _2}({b_e}),
\end{eqnarray*}
and,
\begin{eqnarray*}\label{eq:drpara}
\eta  \equiv {{{T_\parallel }} \over {{T_ \bot }}},~\tau \equiv
{{{T_{\parallel i}}} \over {{T_{\parallel e}}}},{\rm{ }}{\tau _ \bot
} \equiv {{{T_{ \bot i}}} \over {{T_{ \bot e}}}},~\beta  \equiv {{8\pi {n_0}T} \over {B_0^2}},\\
b \equiv {{k_ \bot
^2{\rho ^2}} \over 2},{\rm{  }}{\bar \omega ^2} \equiv
{{{\omega ^2}} \over {k_\parallel ^2v_A^2}},\\
\delta {\phi _\parallel } \equiv \delta \phi  - \delta \psi
~({E_\parallel } =  - i{k_\parallel }\delta {\phi _\parallel
}),\\
{\Gamma _0}(b) = {I_0}(b){e^{ - b}},\\
{\Gamma _1}(b) = \left[
{{I_0}(b) - {I_1}(b)} \right]{e^{ - b}},{\rm{ }}{\Gamma _2}(b) =
2{\Gamma _1}(b).
\end{eqnarray*}
where ${I_n}$  is the first kind modified Bessel function. The
dispersion relation is
\begin{equation}\label{eq:dr}
\det \left|
{\mathord{\buildrel{\lower3pt\hbox{$\scriptscriptstyle\leftrightarrow$}}\over
 C} } \right| = 0
\end{equation}

We see from (\ref{eq:drmatrix}), the matrix is symmetrical, i.e.,
$c_{AS}=c_{SA}$, $c_{AM}=c_{MA}$ and $c_{MS}=c_{SM}$, which is
similar with the full-kinetic one (see e.g., \cite{Stix1992}). When
isotopic ($\eta_{i,e}=1$), (\ref{eq:dr}) reduces to exactly the same
result as \cite{Howes2006}. At appendix, we also show that when drop
 small terms from reduced full-kinetic result in \cite{Yoon1993},
it also gives the same result as this gyrokinetic one.

We can see from (\ref{eq:drmatrix}), the solutions for
$\bar{\omega}$ is independent of ${k_\parallel }$, which means
always ${\omega ^2} \propto k_\parallel ^2$. This is the main
drawback of (\ref{eq:dr}) because of that ${k_\parallel }/{k_ \bot
}$ is taken as small term and dropped, then not suitable to discuss
the fine structures of the dispersion relation, e.g., the maximum
growth rate ${\gamma _{\max }}$ of mirror mode. However, one can
also add the small terms when necessary by comparing with the
full-kinetic result (see appendix). For example, \cite{Qu2007} adds
a ${k_\parallel }/{k_ \bot }$ term in integral form when discuss
mirror mode.

\section{Solutions of Firehose and Mirror Mode}\label{sec:fhmm}\suppressfloats
In (\ref{eq:drmatrix}), when only considering one matrix element,
$c_{SS}=0$ gives ions sound wave (ISW, only consider $\delta {\phi
_\parallel }$), $c_{AA}=0$ gives shear Alfven wave (SAW, only
consider $\delta \psi$) and $c_{MM}=0$ gives mirror mode (MM, only
consider $\delta {B_\parallel }$). $c_{\alpha \beta }(\alpha,
\beta=S, A, M)$ is the coupling of each waves.

\subsection{Firehose (KAW) Branch}
As mentioned above, $c_{AA}=0$  represents the firehose and KAW
solution,
\begin{eqnarray}\label{eq:fh1}
{{\bar \omega }^2} = {{{b_i}} \over {{\lambda _5}}}(1 + \Delta ) =
{{{b_i}} \over {\left[ {1 - {\Gamma _0}({b_i})} \right] + {\tau _
\bot }\left[ {1 - {\Gamma _0}({b_e})} \right]}}\nonumber \\
\left\{ {1 + {{{\beta _{ \bot i}}} \over {2{b_i}}}\left( {1 - {\eta
_i}} \right)\left[ {1 - {\Gamma _0}({b_i})} \right] + {{{\beta _{
\bot e}}} \over {2{b_e}}}\left( {1 - {\eta _e}} \right)\left[ {1 -
{\Gamma _0}({b_e})} \right]} \right\}.
\end{eqnarray}
When small Lamor radius ${b_e} \ll {b_i} \ll 1$, ${\Gamma _0}(b)
\simeq 1 - b + 3{b^2}/4 \simeq 1 - b$, (\ref{eq:fh1}) reduces to,
\begin{equation}\label{eq:fh2}
{\bar \omega ^2} = 1 + \Delta  = 1 + {{{\beta _{ \bot i}}} \over
2}\left( {1 - {\eta _i}} \right) + {{{\beta _{ \bot e}}} \over
2}\left( {1 - {\eta _e}} \right),
\end{equation}
which is the classical firehose solution, and when $1 + \Delta  <
0$, this mode is unstable. When isotropic ${\eta _{i,e}} = 1$,
$\Delta  = 0$, (\ref{eq:fh1}) reduces to
\begin{equation}\label{eq:fh3}
{\bar \omega ^2} = {{{b_i}} \over {\left[ {1 - {\Gamma _0}({b_i})}
\right] + {\tau _ \bot }\left[ {1 - {\Gamma _0}({b_e})} \right]}},
\end{equation}
Still using the small Lamor radius assumption to go further,
(\ref{eq:fh3}) reduces to
\begin{equation}\label{eq:fh4}
{\bar \omega ^2} = 1 + 3{b_i}/4,
\end{equation}
(\ref{eq:fh4}) is the classical KAW solution \cite{Hasegawa1975,
Hasegawa1976}.

\subsection{Mirror Mode Branch}
At this case, $c_{MM}=0$, gives
\begin{equation}\label{eq:mm1}
{{2{\eta _i}} \over {{\beta _{ \bot i}}}} - {\lambda _7} - {\lambda
_8} = 0,
\end{equation}
And, dropping the contributions from electrons (cold electrons
assumption) in (\ref{eq:mm1}), gives,
\begin{equation}\label{eq:mm2}
{\xi _i}Z({\xi _i}) = {{{\eta _i} - (1 - {\eta _i}){\Gamma
_1}({b_i}){\beta _{ \bot i}}} \over {{\beta _{ \bot i}}{\Gamma
_1}({b_i})}}.
\end{equation}
(\ref{eq:mm2}) is the mirror mode solution. \cite{Qu2007} adds a
${k_\parallel }/{k_ \bot }$ term to (\ref{eq:mm2}), then can be used
to discuss the ${\gamma _{\max }}$ of mirror mode, which give the
same result as \cite{Pokhotelov2004} from full-kinetic theory.

\section{Coupling of Firehose and Mirror Mode}\label{sec:coupling}\suppressfloats
In the above section, we use only one matrix element of the 3-by-3
general dispersion matrix (\ref{eq:drmatrix}), then get kinetic
version of both FH (KAW) and MM, which shows accurately enough to
discuss single mode by compared with full-kinetic results from
previous authors.

At this section, we solve the 2-by-2 matrix to discuss coupling
effects of FH and MM. The matrix (\ref{eq:drmatrix}) is general for
arbitrary $b_{i,e}$,
 $\beta_{i,e}$, ${\xi _{i,e}}$ under the gyrokinetic ordering. When cold electrons
($T_e/T_i \ll 1$), we have $b_e \ll 1$, $\tau, {\tau _ \bot } \gg
1$, ${c_{SS}} \gg {c_{SA}},{c_{SM}}$, then $\delta {\phi _\parallel
}$ should be very small, which means the parallel electrical field
is short circuited (few percents of cold electrons in density is
enough, see \cite{Hasegawa1975a}). We use this only assumption to
reduce the 3-by-3 matrix to 2-by-2 for discussion FH and MM
coupling.

(\ref{eq:dr}) reduces to,
\begin{equation}\label{eq:fhmm1}
\left| \begin{array}{cc}
{{{\eta _i}{b_i}} \over {{{\bar \omega }^2}}}(1 + \Delta ) - {\eta _i}{\lambda _5} & {{\eta _i}{\lambda _6}}  \\
{{\eta _i}{\lambda _6}} & {{{2{\eta _i}} \over {{\beta _{ \bot i}}}}
- {\lambda _7} - {\lambda _8}}
\end{array} \right| = 0,
\end{equation}
 where,
\begin{eqnarray}\label{eq:fhmm2}
\Delta  = {{{\beta _{ \bot i}}} \over {2{b_i}}}\left( {1 - {\eta
_i}} \right)\left[ {1 - {\Gamma _0}({b_i})} \right] + {{{\beta _{
\bot e}}} \over 2}\left( {1 - {\eta _e}} \right)  \nonumber \\
\simeq {{{\beta _{ \bot i}}} \over 2}\left( {1 - {\eta _i}}
\right)(1 - {3 \over
4}{b_i}), \nonumber \\
{\lambda _5} \simeq 1 - {\Gamma _0}({b_i})
\simeq {b_i} - {3 \over 4}b_i^2, \nonumber \\
{\lambda _6} \simeq {\Gamma _1}({b_i}) - 1 \simeq  - {3 \over
2}{b_i} + {5 \over 4}b_i^2, \nonumber \\
{\lambda _7} \simeq 2(1 - {\eta _i}){\Gamma _1}({b_i}), \nonumber \\
{\lambda _8} = 2{\xi _i}Z({\xi _i}){\Gamma _1}({b_i}) + {{2{\eta
_i}} \over {{\eta _e}{\tau _ \bot }}}{\xi _e}Z({\xi _e}){\Gamma
_1}({b_e})  \nonumber \\ \simeq 2{\xi _i}Z({\xi _i}){\Gamma
_1}({b_i}).
\end{eqnarray}
The small $b_i$ expansion is also written in (\ref{eq:fhmm2}) but
only used when necessary.

\cite{Pokhotelov2004} has given a very similar 2-by-2 matrix as
(\ref{eq:fhmm1}) via full-kinetic theory, which contains also the
small ${k_\parallel }/{k_ \bot }$ correction. But, it solves only MM
by dropping the firehose correction in the mirror approximation
$\left| \omega  \right| \ll \left| {{k_\parallel }{v_{i,\parallel
}}} \right|$. A fluid theory is used to discuss the instability of
mirror mode and firehose in \cite{Duhau1985}, and several stable and
unstable regions are given in different parameters space. Since
there is no kinetic effect, the results of \cite{Duhau1985} will be
unreasonable. Ignoring FLR effects, \cite{Klimushkin2012} gives the
gyrokinetic treatment of the stability of coupled Alfv\'en and drift
mirror modes in 1D non-uniform plasma.

For analytical tractable, we solve (\ref{eq:fhmm1}) in long
wavelength approximation in this section. This means $b_i \ll 1$.
Keep $O(1) + O(b)$, (\ref{eq:fhmm1}) changes to
\begin{eqnarray}\label{eq:longwave1}
-\left\{ {{{{\eta _i}} \over {{\beta _{ \bot i}}}} - \left[ {(1 -
{\eta _i}) + {\xi _i}Z({\xi _i})} \right]} \right\}\left[ {{1 \over
{{{\bar \omega }^2}}}(1 + \Delta ) - 1} \right]  \cr = \left\{
\begin{array}{c}{ 3
\over 4}{{{\eta _i}} \over {{\beta _{ \bot i}}}} - {3 \over 4}\left[
{(1 - {\eta _i}) + {\xi _i}Z({\xi _i})} \right] +  \\{3 \over
2}\left[ {(1 - {\eta _i}) + {\xi _i}Z({\xi _i})} \right]\left[ {{1
\over {{{\bar \omega }^2}}}(1 + \Delta ) - 1} \right] - {9 \over
8}{\eta _i} \end{array}  \right\}{b_i}
\end{eqnarray}

We use (\ref{eq:longwave1}) to see the correction from MM to FH and
FH to MM.

\subsection{Mirror mode correction to Firehose (KAW)}
Look the small FLR correction to firehose solution $\left[ {(1 +
\Delta )/{{\bar \omega }^2} - 1} \right] \sim O({b_i})$,
\begin{equation}\label{eq:mm2fh1}
{\bar \omega ^2} = {{(1 + \Delta )} \over {1 - {{\left\{ {{3 \over
4}{{{\eta _i}} \over {{\beta _{ \bot i}}}} - {3 \over 4}\left[ {(1 -
{\eta _i}) + {\xi _i}Z({\xi _i})} \right] - {9 \over 8}{\eta _i}}
\right\}{b_i}} \over {\left\{ {{{{\eta _i}} \over {{\beta _{ \bot
i}}}} - \left[ {(1 - {\eta _i}) + {\xi _i}Z({\xi _i})} \right]}
\right\}}}}}.
\end{equation}

\subsubsection{Cold ions}
When cold ions, $\left| \omega  \right| \gg \left| {{k_\parallel
}{v_{i,\parallel }}} \right|$, ${\xi _i} = \bar \omega /\sqrt
{{\beta _{\parallel i}}}  \gg 1$, ${\xi _i}Z({\xi _i}) \simeq  - 1 -
1/({2\xi _i^2}) \simeq  - 1$,

\begin{eqnarray}\label{eq:mm2fhcold}
&{{\bar \omega }^2} = {{(1 + \Delta )} \over {1 - {{{3 \over
8}\left\{ {{2 \over {{\beta _{ \bot i}}}} - 1} \right\}{b_i}}
\mathord{\left/ {\vphantom {{{3 \over 8}\left\{ {{2 \over {{\beta _{
\bot i}}}} - 1} \right\}{b_i}} {\left\{ {{1 \over {{\beta _{ \bot
i}}}} + 1} \right\}}}} \right.
 \kern-\nulldelimiterspace} {\left\{ {{1 \over {{\beta _{ \bot i}}}} + 1}
 \right\}}}}} \nonumber \\
& \buildrel {{\eta _i} = 1,~{\rm{  }}{\beta _{ \bot i}} \ll 1} \over
 \Longrightarrow {{\bar \omega }^2} \simeq 1 + {3 \over 4}{b_i}~{{\rm{(KAW)}}}
\end{eqnarray}

\subsubsection{Hot ions}
Here, we try to see that, when classical FH (KAW) is stable ( $1 +
\Delta  > 0$),  whether MM can excite it to unstable. When hot ions,
$\left| \omega  \right| \ll \left| {{k_\parallel }{v_{i,\parallel
}}} \right|$, ${\xi _i} = \bar \omega /\sqrt {{\beta _{\parallel
i}}}  \ll 1$, ${\xi _i}Z({\xi _i}) \simeq i\sqrt \pi  {\xi _i} =
\bar \omega {{i\sqrt \pi  } \mathord{\left/ {\vphantom {{i\sqrt \pi
} {\sqrt {{\beta _{\parallel i}}} }}} \right.} {\sqrt {{\beta
_{\parallel i}}} }} \simeq {{i\sqrt \pi  \sqrt {1 + \Delta }
{\mathop{\rm sgn}} (\bar \omega )} \mathord{\left/ {\vphantom
{{i\sqrt \pi  \sqrt {1 + \Delta } {\mathop{\rm sgn}} (\bar \omega )}
{\sqrt {{\beta _{\parallel i}}} }}} \right.} {\sqrt {{\beta
_{\parallel i}}} }}$, (Note: in fact, $\bar \omega $  is complex)
\begin{eqnarray}\label{eq:mm2fhhot}
{{\bar \omega }^2} &= {{(1 + \Delta )\left\{
\begin{array}{c} \left[
{{A^2} - {{3{b_i}} \over 4}(A - {3 \over 2}{\eta _i})A + (1 -
{{3{b_i}} \over 4}){B^2}} \right] \\
- \left[ {{9 \over 8}{b_i}{\eta _i}B} \right]i \end{array} \right\}}
\over {{{\left[ {A - {{3{b_i}} \over 4}(A - {3 \over 2}{\eta _i})}
\right]}^2} + {{\left[ {(1 - {{3{b_i}} \over 4})B} \right]}^2}}} \cr
&= a - i{\mathop{sgn}} (\bar \omega )b~\left( {a\sim O(1),~b\sim
O({b_i})} \right)
\end{eqnarray}
where,
\begin{equation}\label{eq:mm2fhhot1}
A \equiv {{{\eta _i}} \over {{\beta _{ \bot i}}}} - (1 - {\eta
_i}),{\rm{ }}B \equiv {{\sqrt \pi  \sqrt {1 + \Delta } {\mathop{\rm
sgn}} (\bar \omega )} \over {\sqrt {{\beta _{\parallel i}}} }}
\end{equation}

Due to $ - {\mathop{\rm sgn}} (\bar \omega )$, and provided, $a>0$,
the solutions are always damped.

To find a unstable mode, which means $a<0$,
\begin{equation}\label{eq:mm2fhhot2}
a < 0 \Leftrightarrow \underbrace {(1 - {{3{b_i}} \over 4}){A^2}}_{
> 0} + \underbrace {{9 \over 8}{b_i}{\eta _i}A}_{ < 0} + \underbrace
{(1 - {{3{b_i}} \over 4}){B^2}}_{ > 0} < 0,
\end{equation}
(\ref{eq:mm2fhhot2}) can only be satisfied when $A<0$, i.e., mirror
mode is unstable. While, if there is a domain satisfy both $a<0$ and
$A<0$, there must be at least one point satisfy both $a=0$ and
$A=0$,
\begin{eqnarray}\label{eq:mm2fhhot3}
\left\{ \begin{array}{c}
  a = 0, \hfill \cr
  A = 0, \hfill \cr \end{array}  \right. \Leftrightarrow \left\{ \begin{array}{c}
  (1 - {{3{b_i}} \over 4}){B^2} = 0, \hfill \cr
  A = 0, \hfill \cr \end{array} \right. \Leftrightarrow \left\{ \begin{array}{c}
  B = 0, \hfill \cr
  A = 0, \hfill \cr \end{array}  \right. \nonumber \\
  \Leftrightarrow \left\{ \begin{array}{c}
  1 + \Delta  = 0, \hfill \cr
  {{{\eta _i}} \over {{\beta _{ \bot i}}}} - (1 - {\eta _i}) = 0, \hfill \cr \end{array} \right.
\end{eqnarray}
(\ref{eq:mm2fhhot3}) can never be satisfied, then, there is no
unstable domain, providing we use small argument expansion of
${\Gamma _{0,1}}({b_i})$  and $Z({\xi _i})$.

\begin{figure}[!htbp]
\centering
\includegraphics[width=8.5cm]{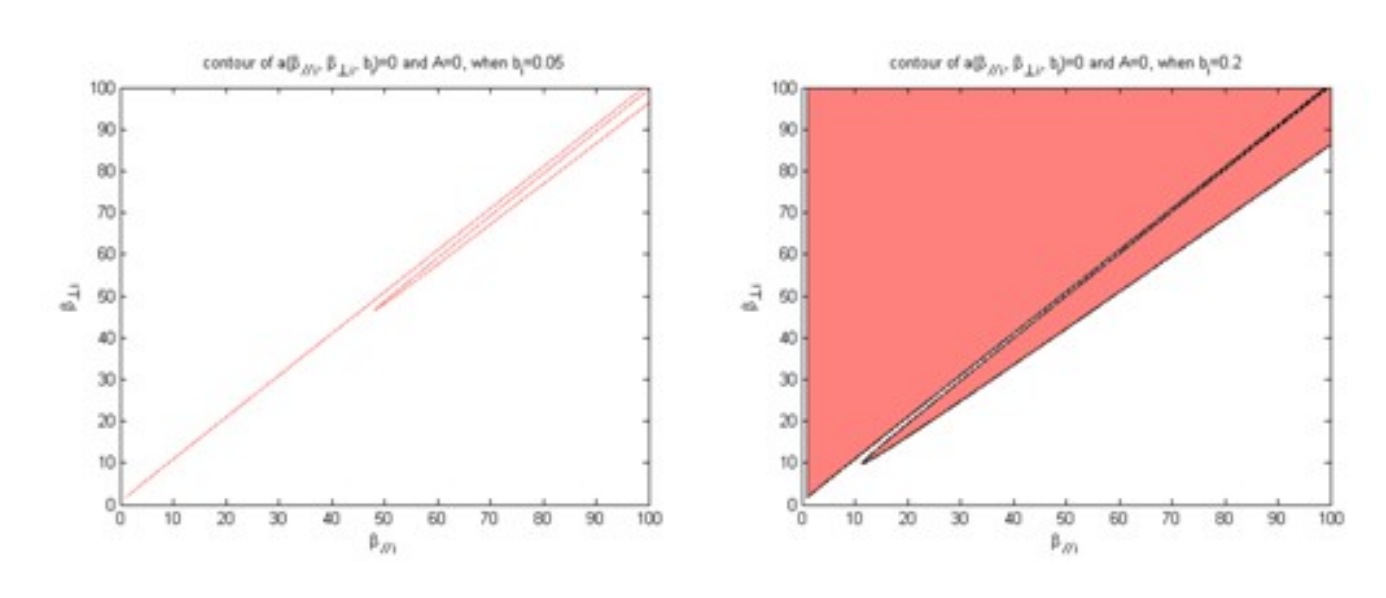}
\caption{\label{fig1} Contour plot of $a=0$ and $A=0$.}
\end{figure}

We can also contour plot (see FIG.\ref{fig1}) the $a=0$ boundary and
$A=0$ boundary to show whether there is a region satisfy both
(\ref{eq:mm2fhhot2}) and $A<0$. However, these contour plots can
also help to show the sizes of each kind of ranges.

We see that, indeed, none of the parameters $({\beta _{\parallel
i}},~{\beta _{ \bot i}})$ satisfy both $a<0$ and $A<0$, which means
that the unstable domain, if it exists, should be very very narrow.
The only possibility is when ${\Gamma _{0,1}}({b_i})$ or $Z({\xi
_i})$ cannot be expanded, i.e., ${b_i}\sim 1$ or ${\xi _i}\sim 1$.
In next section, we will use Nyquist technique to prove that there
is indeed no unstable domain for exciting the classical stable FH
(KAW) in arbitrary parameters.

\subsubsection{Numerical solutions}
Numerical solutions of (\ref{eq:fhmm1}) at FH (KAW) branch are given below,

\begin{figure}[!htbp]
\centering
\includegraphics[width=8.5cm]{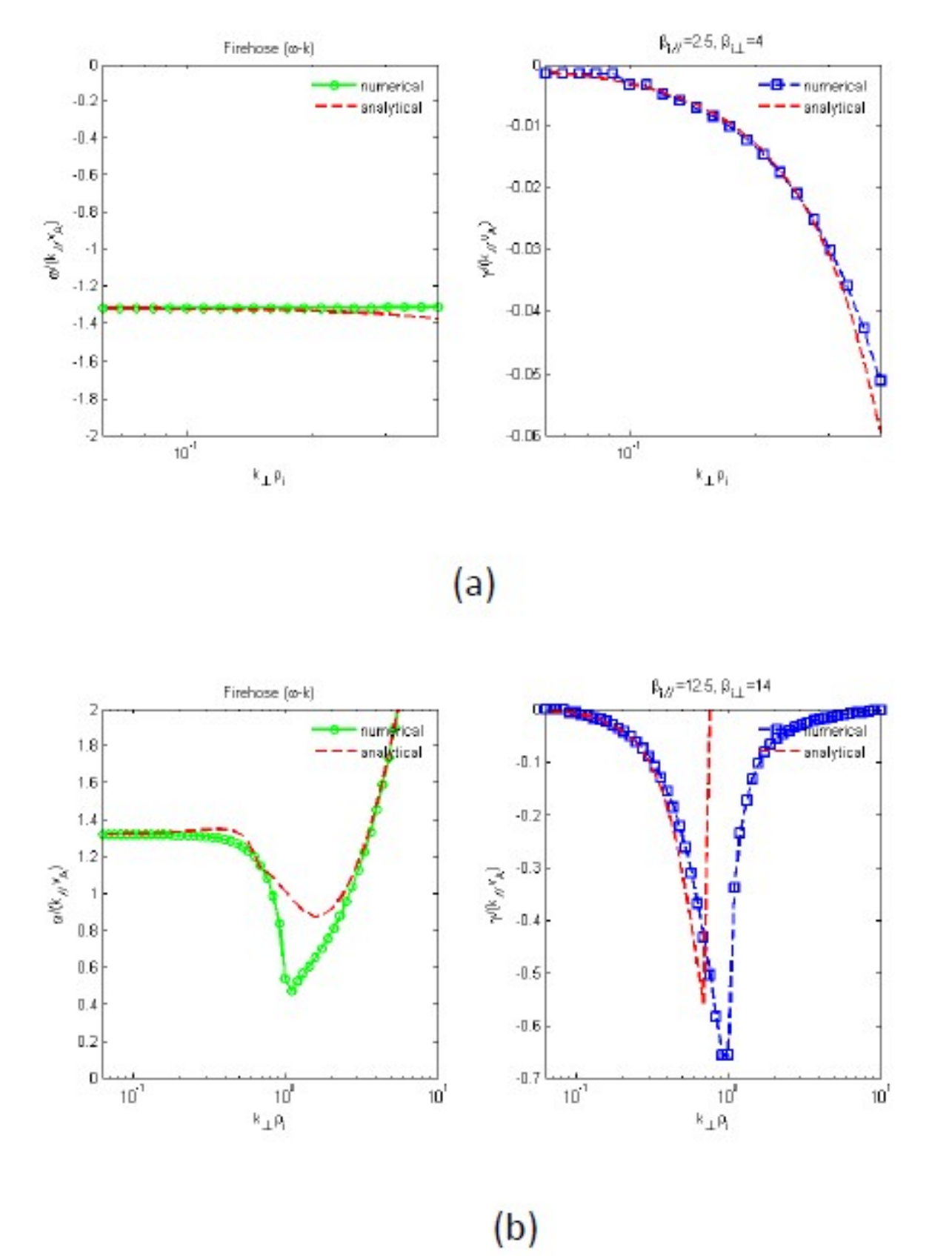}
\caption{\label{fig2} Numerical solution and analytical solution
(\ref{eq:mm2fhhot}), firehose (KAW).}
\end{figure}

FIG.\ref{fig2} shows, the analytical and numerical solutions are
agreed very well. The coupling of MM is weak, and mainly brings KAW
(FH) a damping.

\subsection{Firehose correction to mirror mode}
At this case,
\begin{equation}\label{eq:fh2mm1}
\left[ {{{2{\eta _i}} \over {{\beta _{ \bot i}}}} - {\lambda _7} -
{\lambda _8}} \right] = {\eta _i}\lambda _6^2/\left[ {{{{b_i}} \over
{{{\bar \omega }^2}}}(1 + \Delta ) - {\lambda _5}} \right],
\end{equation}
We define ${\bar \omega _0}$ be the traditional mirror mode solution
(zeroth order), i.e., ${\xi _{i0}} = {\bar \omega _0}/\sqrt {{\beta
_{\parallel i}}}  \ll 1$, and
\begin{equation}\label{eq:fh2mm2}
{\xi _{i0}}Z({\xi _{i0}}) = {{{\eta _i} - (1 - {\eta _i}){\Gamma
_1}({b_i}){\beta _{ \bot i}}} \over {{\beta _{ \bot i}}{\Gamma
_1}({b_i})}},
\end{equation}

\begin{figure}[!htbp]
\centering
\includegraphics[width=8.5cm]{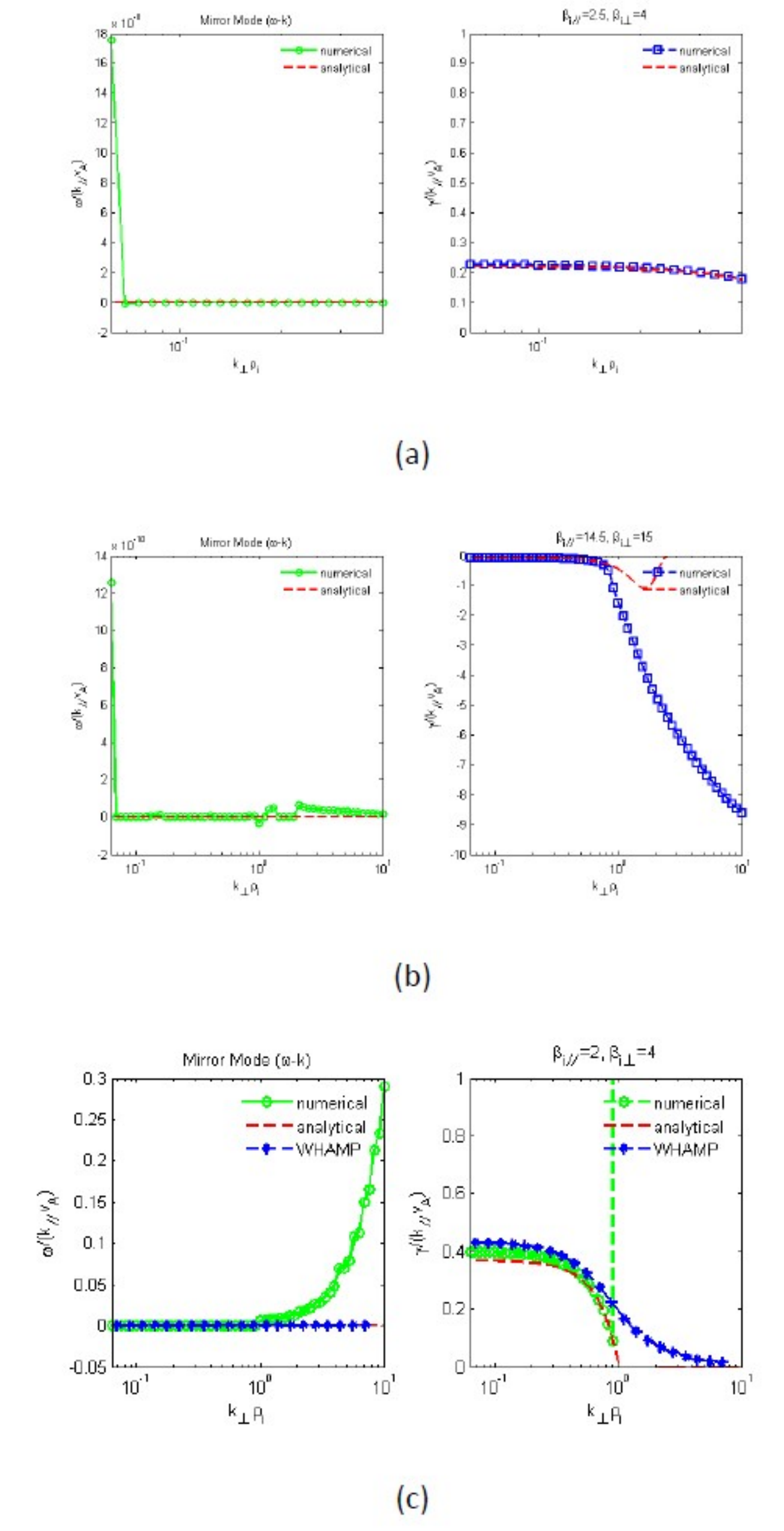}
\caption{\label{fig3} Numerical solution of (\ref{eq:fhmm1}) and analytical
solution from (\ref{eq:fh2mm3}), mirror mode.}
\end{figure}

We can rewrite (\ref{eq:fh2mm1}) as,
\begin{equation}\label{eq:fh2mm3}
{\xi _i}Z({\xi _i}) \simeq \left[ {{{{\eta _i}} \over {{\beta _{
\bot i}}}} - (1 - {\eta _i})} \right] - {{{\eta _i}{{[{\Gamma
_1}({b_i}) - 1]}^2}} \over {2\left\{ {{{{b_i}} \over {\bar \omega
_0^2}}(1 + \Delta ) - \left[ {1 - {\Gamma _0}({b_i})} \right]}
\right\}}},
\end{equation}
which brings just a very small correction and MM is still pure
growing or damped. This means the coupling from KAW (FH) is also
weak, and the traditional treatment of MM reasonable. To go further,
we use the expansion ${\xi _i}Z({\xi _i}) \simeq i{\xi _i}\sqrt \pi
\left[ {{e^{ - \xi _i^2}} - 2{\xi _i}} \right] \simeq i\sqrt \pi
\bar \omega /\sqrt {{\beta _{\parallel i}}} $. The analytical
solution in FIG.\ref{fig3} is by solving (\ref{eq:fh2mm2}) and
(\ref{eq:fh2mm3}) with ${\xi _i}Z({\xi _i}) \simeq i\sqrt \pi  \bar
\omega /\sqrt {{\beta _{\parallel i}}} $, and then adding the
correction term $\left[ {{e^{ - \xi _i^2}} - 2{\xi _i}} \right]$.
Numerical solutions of (\ref{eq:fhmm1}) at mirror mode branch are
also given.

From FIG.\ref{fig3}(b), we can see, for mirror mode, the analytical
solution is only suitable when ${b_i} \ll 1$, because, when ${b_i} >
\sim 1$, ${\xi _i} = \bar \omega /\sqrt {{\beta _{\parallel i}}}
\sim 1$ . At FIG.\ref{fig3}(c), the mirror mode assumption is
totally broken when $b_i$ is large. WHAMP result is also shown.

\subsection{Physics Mechanism}
\subsubsection{Mirror mode to firehose (KAW)}
We find from the analytical and numerical solution above, that, the
correction from mirror mode to firehose (KAW) is always bring not a
unstable but a damping. This seems mean that the mirror cannot
excite firehose but, on the contrary, absorb energy from firehose.

It's widely accepted that the mechanism of mirror mode unstable is
both the Landau type (wave-particle resonant) and the anisotropic
free energy type. Then, it is physical reasonable that when mirror
mode is unstable, it absorbs energy from firehose and cause which
damped.

\subsubsection{Firehose to mirror mode}
Mirror mode is always pure growing or damping (${\mathop{\rm
Re}\nolimits} (\left| {\bar \omega } \right|) \ll {\mathop{\rm
Im}\nolimits} (\left| {\bar \omega } \right|)$), even though with
firehose coupling. There is a narrow range (${\beta _{ \bot i}} >
\sim {\beta _{\parallel i}}$), both mirror mode and firehose are
damping, which may be caused by Landau damping, similar with
beam-plasma system (a range the free energy of non-Maxwellian
distribution cannot compete to Landau damping).

\section{Nyquist Stabilities Analysis}\label{sec:nyquist}\suppressfloats
The dispersion relation (\ref{eq:fhmm1}) can be rewritten as
\begin{equation}\label{eq:nyquist1}
D(\omega ) = ({\bar \omega ^2} - {\alpha _0})({\alpha _1} - {\xi
_i}Z({\xi _i})) + {\bar \omega ^2}{c^2} = 0,
\end{equation}
where
\begin{eqnarray}\label{eq:nyquist2}
{\alpha _0} = {{{b_i}} \over {1 - {\Gamma _0}({b_i})}}(1 + \Delta
),{\rm{ }}{\alpha _1} = {{{\eta _i}} \over {{\beta _{ \bot
i}}{\Gamma _1}({b_i})}} - (1 - {\eta _i}), \nonumber \\
{\rm{  }}{c^2} = {{{\eta _i}{{[{\Gamma _1}({b_i}) - 1]}^2}} \over
{2{\Gamma _1}({b_i})[1 - {\Gamma _0}({b_i})]}}.
\end{eqnarray}

Classical results of firehose and mirror mode instabilities,
\begin{eqnarray}\label{eq:nyquist3}
{\rm{Firehose:}}~\left\{ \begin{array}{c}
  {\rm{stable:}}~{\alpha _0} > 0, \hfill \cr
  {\rm{unstable:}}~{\alpha _0} < 0. \hfill \cr \end{array}  \right.; \nonumber \\
  {\rm{Mirror mode:}}~\left\{ \begin{array}{c}
  {\rm{stable:}}~{\alpha _1} > 0, \hfill \cr
  {\rm{unstable:}}~{\alpha _1} < 0. \hfill \cr\end{array}  \right..
\end{eqnarray}

For different $({\beta _\parallel },{\rm{ }}{\beta _ \bot })$, there
are three regions: (1) ${\alpha _0} > 0,{\rm{ }}{\alpha _1} < 0$,
mirror mode unstable region; (2) ${\alpha _0} < 0,{\rm{ }}{\alpha
_1} > 0$, firehose unstable region; (3) ${\alpha _0} > 0,{\rm{
}}{\alpha _1} > 0$, both stable (damping) region.

\subsection{Mirror mode unstable region}
\begin{figure}[!htbp]
\centering
\includegraphics[width=8.5cm]{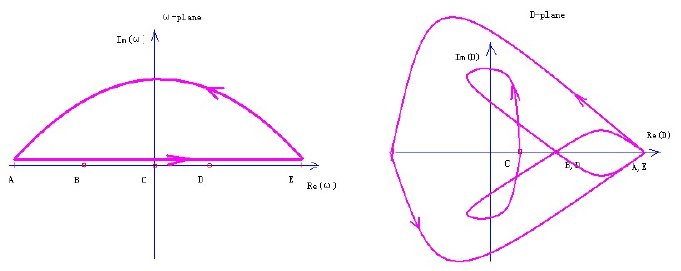}
\caption{\label{fig4} Nyquist diagram of (\ref{eq:nyquist1}), mirror
mode unstable region.}
\end{figure}

We use Nyquist technique to prove that when ${\alpha _0} > 0,{\rm{
}}{\alpha _1} < 0$(mirror mode unstable region), there is only one
unstable mode (i.e., mirror mode).
\begin{equation}\label{eq:nyquistmm1}
{\xi _i}Z({\xi _i}) = {\xi _i}{Z_{ir}} + i{\xi _i}{Z_{ii}},
\end{equation}
\begin{equation}\label{eq:nyquistmm2}
D(\omega ) = ({\bar \omega ^2} - {\alpha _0})({\alpha _1} - {\xi
_i}{Z_{ir}}) + {\bar \omega ^2}{c^2} - i({\bar \omega ^2} - {\alpha
_0}){\xi _i}{Z_{ii}},
\end{equation}
When $\left| \omega  \right| \to \infty $,
\begin{equation}\label{eq:nyquistmm3}
D(\omega ) \to {\bar \omega ^2}({\alpha _1} + 1 + {c^2}) > 0,
\end{equation}
Along  $ - \infty  < {\omega _r} < \infty$, at ${\mathop{\rm
Im}\nolimits} D(\omega ) = 0$
\begin{equation}\label{eq:nyquistmm4}
\left\{ \begin{array}{c}
  {{\bar \omega }^2} = {\alpha _0},{\rm{  }}D = {\alpha _0}{c^2} > 0, \hfill \cr
  \bar \omega  = 0,{\rm{   }}D =  - {\alpha _0}{\alpha _1} > 0. \hfill \cr \end{array}  \right.
\end{equation}
The Nyquist diagram will draw as FIG.\ref{fig4}.

We see from FIG.\ref{fig4}, the mapping in complex D-plane indeed
encircle the origin point, and one time, which means there is indeed
only one unstable mode. A numerical result of the mapping (partly)
is shown below in FIG.\ref{fig5}

\begin{figure}[!htbp]
\centering
\includegraphics[width=8.5cm]{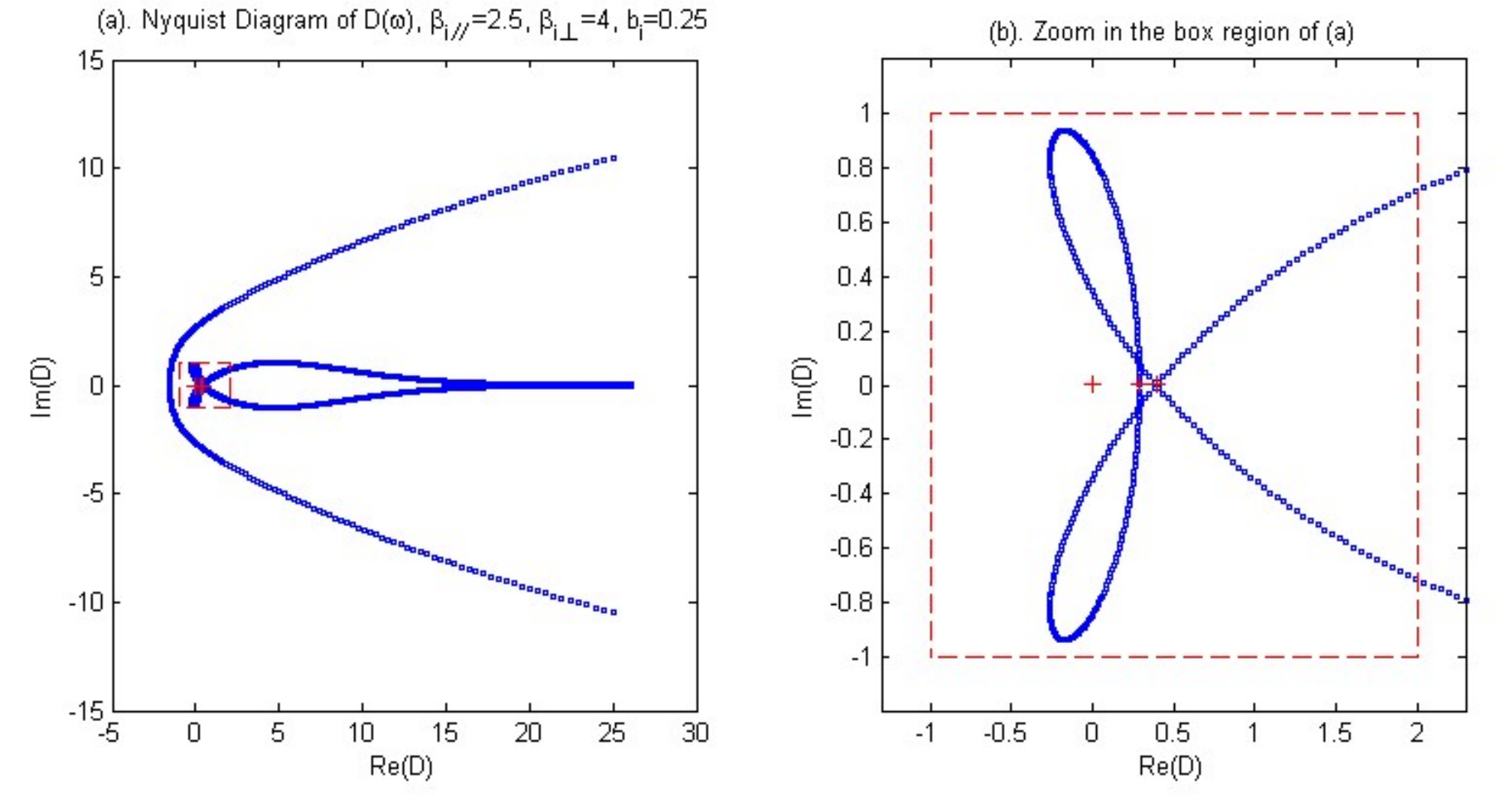}
\caption{\label{fig5} Nyquist diagram of (\ref{eq:nyquist1})
verified by numerical, mirror mode unstable region.}
\end{figure}

\subsection{Firehose unstable region}
\begin{figure}[!htbp]
\centering
\includegraphics[width=8.5cm]{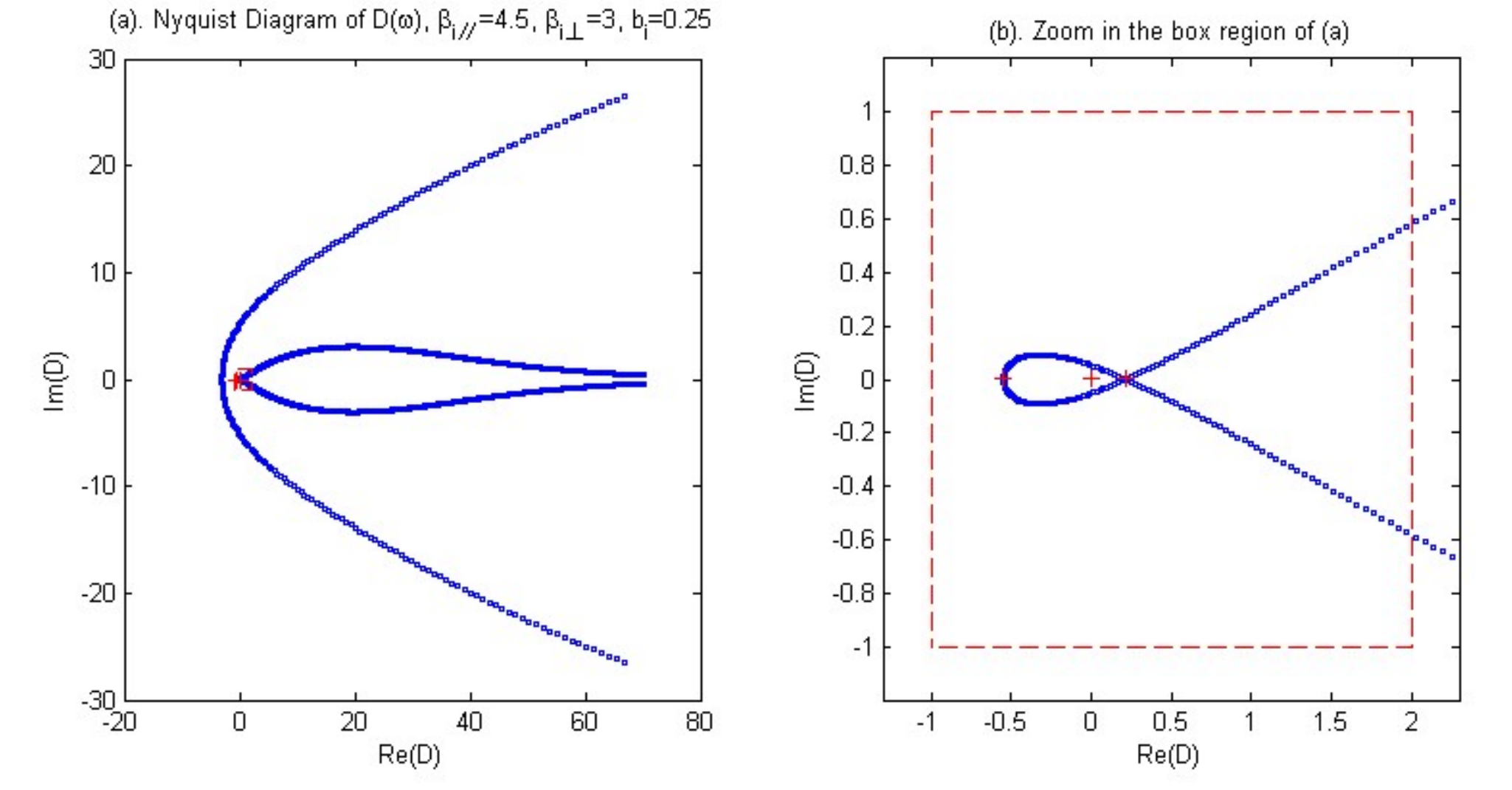}
\caption{\label{fig6} Nyquist diagram of (\ref{eq:nyquist1})
verified by numerical, firehose unstable region.}
\end{figure}

At this case ${\alpha _0} < 0,{\rm{ }}{\alpha _1} > 0$. When $\left|
\omega  \right| \to \infty $, we still have (\ref{eq:nyquistmm3}).
And, along $ - \infty  < {\omega _r} < \infty$, ${\mathop{\rm
Im}\nolimits} D(\omega ) = 0$ at only $\bar \omega  = 0$, which
gives $D =  - {\alpha _0}{\alpha _1} > 0$. Then, after drawing the
Nyquist diagram, we can also get only one unstable mode (i.e.,
firehose). A result is shown in FIG.\ref{fig6}

\subsection{Both stable (damping) region}
\begin{figure}[!htbp]
\centering
\includegraphics[width=8.5cm]{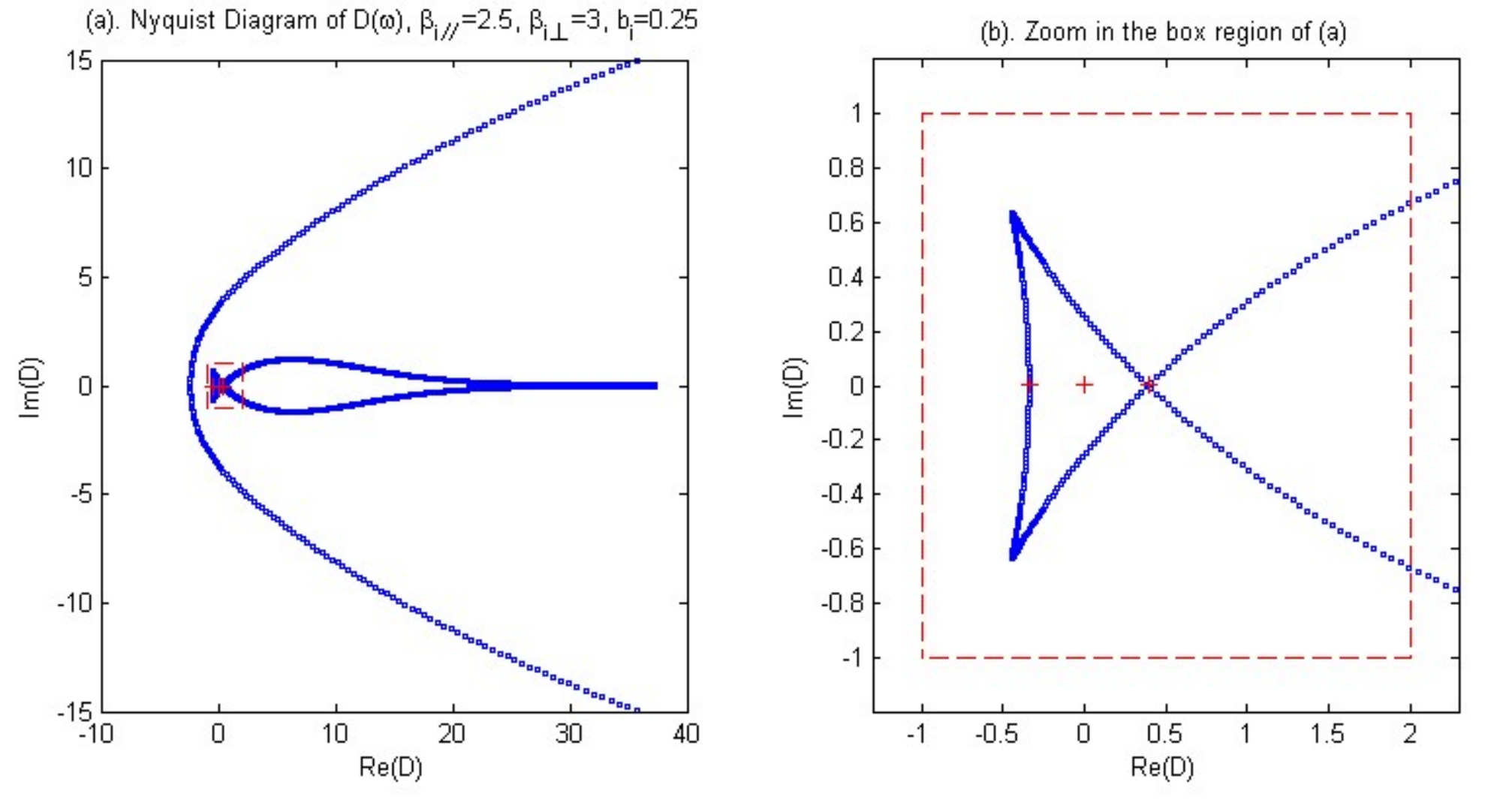}
\caption{\label{fig7} Nyquist diagram of (\ref{eq:nyquist1})
verified by numerical, stable region.}
\end{figure}

At this case ${\alpha _0} > 0,{\rm{ }}{\alpha _1} > 0$, the analysis
is similar with the mirror mode unstable case, except that in Fig4,
the point $C>0$ in $D$-plane should be changed to $C<0$. Then, the
mapping in complex D-plane will never encircle the origin point,
which means all the solutions are stable. A result is shown in
FIG.\ref{fig7}

\section{Summary}\suppressfloats
In this paper, we firstly give a general 3-by-3 gyrokinetic
dispersion matrix for anisotropic bi-Maxwellain distribution plasma
to discuss low-frequency hydromagnetic waves, which is suitable for
arbitrary $b_{i,e}$, $\beta_{i,e}$, $\xi_{i,e}$ under the
gyrokinetic ordering. This gyrokinetic matrix is useful to discuss
various kinetic corrections to hydrodynamica waves. Some detailed
solutions are given in \cite{Howes2006} for isotropic plasma, e.g.,
analytical form of high $\beta$ KAW. However, the gyrokinetic matrix
can also be extended to arbitrary distribution and arbitrary
species, and also to nonuniform plasma. One can do this by following
the framework in \cite{Chen1991}, see e.g., \cite{Klimushkin2012}
for 1D nonuniform plasma case. As an application example, we use
this dispersion matrix to discuss the coupling of firehose and
mirror mode. At the cold electron assumption, the matrix is reduced
to 2-by-2, for the coupling of firehose (KAW) and mirror mode. The
dispersion relation is solved both analytical and numerical, and
shows consistent very well between each other. The results by the
kinetic code WHAMP with all kinetic effects also show the reduced
dispersion relation is appropriate. We find there exists at most one
unstable solution at the $({\beta _\parallel },{\rm{ }}{\beta _ \bot
})$ plane for arbitrary Lamor radius, and the plane can be divided
into three regions: firehose unstable region; mirror mode unstable
region; both modes stable (damping) region. The Nyquist analysis
confirms these three regions.

\section{Acknowledgments}\suppressfloats
One author, HSX, thanks the discussions with Ling CHEN, Hong-peng QU
and Peter. H. YOON, and also thanks Richard E. DENTON for providing
the java version of WHAMP.

\section{Appendix}\label{sec:appendix}\suppressfloats
At this appendix, we compare the full-kinetic dispersion relation
from \cite{Yoon1993} with our gyrokinetic one.

Drop the electrons term as \cite{Yoon1993} by assuming $b_e \ll 1$,
(\ref{eq:dr}) can be rewritten to,
\begin{eqnarray}\label{eq:appendix1}
[{\eta _i}DAC + {\eta _i}{\eta _i}{E^2}A - {\eta _i}D{B^2}]{\bar
\omega ^2} + {\eta _i}{b_i}(1 + \Delta ) \nonumber \\
\left[ {C({\eta
_i}D - A) + {{({\eta _i}E - B)}^2}} \right] = 0,
\end{eqnarray}
where,
\begin{eqnarray}\label{eq:appendix2}
\Delta  = {{{\beta _{ \bot i}}} \over {2{b_i}}}\left( {1 - {\eta
_i}} \right)\left[ {1 - {\Gamma _0}({b_i})} \right] + {{{\beta _{
\bot e}}} \over 2}\left( {1 - {\eta _e}} \right), \cr A = {{Z'({\xi
_i})} \over 2}{\Gamma _0}({b_i}) + \tau {{Z'({\xi _e})} \over 2},
\cr B = {{Z'({\xi _i})} \over 2}{\Gamma _1}({b_i}) - {{{\eta _i}}
\over {{\eta _e}}}{{Z'({\xi _e})} \over 2}, \cr C = {{2{\eta _i}}
\over {{\beta _{ \bot i}}}} + 2{\eta _i}{\Gamma _1}({b_i}) +
2{{{\eta _i}} \over {{\tau _ \bot }}} + {{Z'({\xi _i})} \over
2}{\Gamma _2}({b_i}) + {{{\eta _i}} \over {{\eta _e}{\tau _ \bot
}}}{{Z'({\xi _e})} \over 2}, \cr D = \left[ {1 - {\Gamma _0}({b_i})}
\right] = {\lambda _6}, \cr
E = \left[ {1 - {\Gamma _1}({b_i})}
\right] =  - {\lambda _7}.
\end{eqnarray}

Dropping the $O({\delta ^2})$ term $k_z^2v_A^2/\Omega _i^2$ in
\cite{Yoon1993}, keep only $O(1)$ term in $\eta$ and $\eta '$, the
full-kinetic dispersion relation in \cite{Yoon1993} can be written
as \footnote{For application, all $B/2$ should be changed to $B$,
which is typo in [Yoon1993]. This has been confirmed by private
communication with the author Peter H. Yoon.},
\begin{eqnarray}\label{eq:appendix3}
\underbrace { - A\varepsilon '{{({\omega  \over {{\Omega
_i}}})}^4}}_{O({\delta ^4})} + \underbrace {(2A - {b_i}{\eta _i})(1
+ \Delta ){{k_\parallel ^2v_A^2} \over {\Omega _i^2}}{{({\omega
\over {{\Omega _i}}})}^2}}_{O({\delta ^4})} \cr + \underbrace
{{{k_\parallel ^2v_A^2} \over {\Omega _i^2}}(1 + \Delta
){{k_\parallel ^2v_A^2} \over {\Omega _i^2}}(1 + \Delta )\left[
{\left( {{b_i}{\eta _i} - A} \right)} \right]}_{O({\delta ^4})} &\cr
+ \underbrace {\left[ {{b_i}{\eta _i}{{AC} \over {{\eta _i}}} +
A{E^2} - {1 \over {4{\eta _i}}}{b_i}{B^2}} \right]{{({\omega  \over
{{\Omega _i}}})}^2}}_{{b_i}O({\delta ^2})}+  \cr
\underbrace
{{{k_\parallel ^2v_A^2} \over {\Omega _i^2}}(1 + \Delta ){b_i}{\eta
_i}\left[ {{1 \over {{\eta _i}}}C\left( {D - {A \over {{\eta _i}}}}
\right) + {{\left( {E - {B \over {2{\eta _i}}}} \right)}^2}}
\right]}_{{b_i}O({\delta ^2})} = 0
\end{eqnarray}
where,
\begin{equation}\label{eq:appendix4}
\varepsilon ' = {{1 - {I_0}{e^{ - {\lambda _i}}}} \over {{\lambda
_i}}} - 2{\lambda _i}\sum\limits_{n = 1}^\infty  {{{2{{\left(
{{I_n}{e^{ - {\lambda _i}}}} \right)}^\prime }} \over {{n^2}}}} \sim
O(1)
\end{equation}

Dropping the high order $O({\delta ^4})$ terms in
(\ref{eq:appendix3}), we get exactly the same result as the
gyrokinetic result (\ref{eq:appendix1}). We find in
(\ref{eq:appendix3}) when ${b_i}\sim k_\parallel ^2v_A^2/\Omega
_i^2$, the ${b_i}O({\delta ^2})$ will jump to $O({\delta ^4})$!! So,
we cannot drop the $O({\delta ^4})$ in (\ref{eq:appendix3}) when
${b_i} \to 0$. This is what \cite{Yoon1993} found.

In \cite{Yoon1993}, the classical firehose solution is found in
full-kinetic at the $O({\delta ^4})$ terms, because ${b_i}O({\delta
^2})$ terms can be dropped when ${b_i} \to 0 \ll O({\delta ^2})$.
While in gyrokinetic, the solution is found in ${b_i}O({\delta ^2})$
terms, because we have ruled out the $O({\delta ^4})$ terms by
gyrokinetic assumptions. It should be just a coincidence that they
give the same firehose result. The different can also be explained
in another way: to get the classical firehose solution, the
full-kinetic approach of \cite{Yoon1993} uses the assumption ${b_i}
\to 0 \ll O({\delta ^2})$, while the gyrokinetic approach uses
${b_i} \to 0 \gg O({\delta ^2})$.

To discussion the fine structure of the hydrodynamica waves and
instabilities, (\ref{eq:appendix3}) can be seen as an extent version
of our gyrokinetic dispersion relation, and won't go so complicated
that can only be solved via numerical. This is why we indeed have
general dispersion relation for arbitrary distribution and with all
kinetic effects in \cite{Stix1992} and also can be numerical solved
in very general form by WHAMP (\cite{Ronnmark1983}, but should note
that WHAMP is not suitable for strong damped waves, because the
plasma dispersion function in that code is expanded via P\'ade
approximation), but we still derive many other reduced dispersion
relations to meet special desires.


\begin{thebibliography}{99}\suppressfloats

\bibitem{Brizard2007}  Brizard, A. J. and Hahm, T. S., Foundations of
nonlinear gyrokinetic theory, Rev. Mod. Phys., American Physical
Society, 2007, 79, 421-468.

\bibitem{Chen1991}  Chen, L. and Hasegawa, A.,
Kinetic Theory of Geomagnetic Pulsations, 1. Internal Excitations by
Energetic Particles, JOURNAL OF GEOPHYSICAL RESEARCH, AIP, 1991, 96,
1503-1512.

\bibitem{Chandrasekhar1958}  Chandrasekhar, S., Kaufman, A. N.,
and Watson, K. M., The Stability of the Pinch, Proc. Roy. Soc. Ser.
A 245, 435, 1958.

\bibitem{Chen2010}  Chen, L. and Wu, D. J., Kinetic Alfven
wave instability driven by electron temperature anisotropy in
high-beta plasmas, Physics of Plasmas, AIP, 2010, 17, 062107.


\bibitem{Duhau1985}  Duhau, S. and de La Torre, A., Hydromagnetic
waves for a collisionless plasma in strong magnetic fields, Journal
of Plasma Physics, 1985, 34, 67-76.

\bibitem{Hasegawa1969}  Hasegawa, A., Drift Mirror
Instability in the Magnetosphere, Phys. Fluids, 1969, 12, 2642.


\bibitem{Hasegawa1975}  Hasegawa, A. and Chen, L., Kinetic Process
of Plasma Heating Due to Alfv¨¦n Wave Excitation, Phys. Rev. Lett.,
American Physical Society, 1975, 35, 370-373.

\bibitem{Hasegawa1975a}  Hasegawa, A.,
Plasma Instabilities and Nonlinear Effects, Springer, 1975.

\bibitem{Hasegawa1976}  Hasegawa, A. and Chen, L., Kinetic processes
in plasma heating by resonant mode conversion of Alfven wave,
Physics of Fluids, AIP, 1976, 19, 1924-1934.

\bibitem{Howes2006}  Howes et al,
Astrophysical Gyrokinetics: Basic Equations and Linear Theory, The
Astrophysical Journal, 2006, 651, 590.

\bibitem{Klimushkin2011}  Klimushkin,
D. Y. and Mager, P. N., Spatial structure and stability of coupled
Alfv¨¦n and drift compressional modes in non-uniform magnetosphere:
Gyrokinetic treatment, Planetary and Space Science, 2011, 59, 1613 -
1620.

\bibitem{Klimushkin2012}  Klimushkin, D. Y. and Mager, P. N., Coupled
Alfv¨¦n and drift-mirror modes in non-uniform space plasmas: a
gyrokinetic treatment, Plasma Physics and Controlled Fusion, 2012,
54, 015006.

\bibitem{Parker1958}  Parker, E. N., Dynamical Instability in an
Anisotropic Ionized Gas of Low Density, Phys. Rev., American
Physical Society, 1958, 109, 1874-1876.

\bibitem{Pokhotelov2000}
Pokhotelov, O. A.; Balikhin, M. A.; Alleyne, H. S.-C. K. and
Onishchenko, O. G., Mirror instability with finite electron
temperature effects, JOURNAL OF GEOPHYSICAL RESEARCH, 2000, 105,
2393-2401.

\bibitem{Pokhotelov2001}  Pokhotelov, O. A.; Balikhin, M. A.;
Treumann, R. A. and Pavlenko, V. P., Drift mirror instability
revisited, 1, Cold electron temperature limit, JOURNAL OF
GEOPHYSICAL RESEARCH, 2001, 106, 8455-8463.

\bibitem{Pokhotelov2002}
Pokhotelov, O. A.; Treumann, R. A.; Sagdeev, R. Z.; Balikhin, M. A.;
Onishchenko, O. G.; Pavlenko, V. P. and Sandberg, I., Linear theory
of the mirror instability in non-Maxwellian space plasmas, JOURNAL
OF GEOPHYSICAL RESEARCH, 2002, 107, 1312.

\bibitem{Pokhotelov2003}  Pokhotelov,
O. A.; Sandberg, I.; Sagdeev, R. Z.; Treumann, R. A.; Onishchenko,
O. G.; Balikhin, M. A. and Pavlenko, V. P., Slow drift mirror modes in
finite electron-temperature plasma: Hydrodynamic and kinetic drift
mirror instabilities, JOURNAL OF GEOPHYSICAL RESEARCH, 2003, 108,
1098.

\bibitem{Pokhotelov2004}  Pokhotelov, O. A.; Sagdeev, R. Z.; Balikhin,
M. A. and Treumann, R. A., Mirror instability at finite ion-Larmor
radius wavelengths, JOURNAL OF GEOPHYSICAL RESEARCH, 2004, 109,
A09213.

\bibitem{Pokhotelov2006}  Pokhotelov, O.; Sagdeev, R.; Balikhin, M.
and Treumann, R., Mirror instability including finite Larmor radius
effects, Advances in Space Research, 2006, 37, 1550 - 1555.

\bibitem{Qu2007}
Qu, H.; Lin, Z. and Chen, L., Gyrokinetic theory and simulation of
mirror instability, Phys. Plasmas, 2007, 14, 042108.

\bibitem{Qu2008}  Qu,
H.; Lin, Z. and Chen, L., Nonlinear saturation of mirror instability,
GEOPHYSICAL RESEARCH LETTERS, 2008, 35, L10108.

\bibitem{Qu2008a}  Qu, H. and
Lin, Z., Gyrokinetic particle simulation of compressional
electromagnetic modes, Commun. Comput. Phys., 2008, 4, 519-536.

\bibitem{Ronnmark1983}  Ronnmark, K., Computation of the
dielectric tensor of a Maxwellian plasma, Plasma Physics, 1983, 25,
699.

\bibitem{Rosenbluth1956}  Rosenbluth, M. N., Los Alamos
Lab. Rep. LA-2030 (Los Alamos National Laboratory, Los Alamos, New
Mexico, 1956). (Unpublished)

\bibitem{Rudakov1961}  Rudakov, L. I. and Sagdeev, R. Z., On the
instability of nonuniform rarefied plasma in a strong magnetic
field, Dokl. Akad. Nauk SSSR,. Engl. Transl., 1961, 6, 415.

\bibitem{Schekochihin2008}  Schekochihin, A. A.; Cowley, S. C.; Kulsrud, R.
M.; Rosin, M. S. and Heinemann, T., Nonlinear Growth of Firehose and
Mirror Fluctuations in Astrophysical Plasmas, Phys. Rev. Lett.,
American Physical Society, 2008, 100, 081301.

\bibitem{Southwood1993}
Southwood, D. J. and Kivelson, M. G., Mirror Instability:, 1. Physical
Mechanism of Linear Instability, JOURNAL OF GEOPHYSICAL RESEARCH,
1993, 98, 9181-9187.

\bibitem{Stix1992}  Stix, T., Waves in Plasmas, AIP
Press, 1992.

\bibitem{Tajiri1967}  Tajiri, M., Propagation of Hydromagnetic
Waves in Collisionless Plasma. II. Kinetic Approach, Journal of the
Physical Society of Japan, The Physical Society of Japan, 1967, 22,
1482-1494.

\bibitem{Treumann1997}  Treumann, R. A. and Baumjohann, W.,
Advanced Space Plasma Physics, World Scientific, 1997.

\bibitem{Yoon1993}
Yoon, P. H.; Wu, C. S. and de Assis, A. S., Effect of finite ion
gyroradius on the fire-hose instability in a high beta plasma,
Physics of Fluids B: Plasma Physics, AIP, 1993, 5, 1971-1979.

\bibitem{Yoon1995}  Yoon, P. H., Garden-hose instability in high-beta
plasmas, Physica Scripta, 1995, 1995, 127.

\end{thebibliography}
\end{document}